\documentclass[reprint,superscriptaddress,amsmath,amsfonts,amssymb,aps,prl,showkeys, longbibliography]{revtex4-2}
\usepackage{hyperref}
\usepackage{graphicx}
\usepackage{xcolor}
\hypersetup{
  breaklinks = {true},
  citecolor = {blue},
  colorlinks = {true},
  linkcolor = {red},
}
\usepackage{bm}

\begin{document}

\title{Optimal spin-charge interconversion in graphene through spin-pseudospin entanglement control}

\author{Joaqu{\'i}n Medina Due{\~n}as}
\email[Corresponding Author ]{joaquin.medina@icn2.cat}
\affiliation{Institut Català de Nanociència i Nanotecnologia (ICN2), CSIC and BIST, Campus UAB, Bellaterra 08193, Barcelona, Spain}
\affiliation{Department of Physics, Universitat Aut{\`o}noma de Barcelona (UAB), Campus UAB, Bellaterra, 08193 Barcelona, Spain}

\author{Santiago Gim{\'e}nez de Castro}
\affiliation{Institut Català de Nanociència i Nanotecnologia (ICN2), CSIC and BIST, Campus UAB, Bellaterra 08193, Barcelona, Spain}

\author{Jos{\'e} H. Garc{\'i}a}
\email[Corresponding Author ]{josehugo.garcia@icn2.cat}
\affiliation{Institut Català de Nanociència i Nanotecnologia (ICN2), CSIC and BIST, Campus UAB, Bellaterra 08193, Barcelona, Spain}

\author{Stephan Roche}
\affiliation{Institut Català de Nanociència i Nanotecnologia (ICN2), CSIC and BIST,
Campus UAB, Bellaterra 08193, Barcelona, Spain}
\affiliation{ICREA --- Instituci\'o Catalana de Recerca i Estudis Avan\c{c}ats, 08010 Barcelona, Spain}

\date{\today}

\begin{abstract}
The electrical generation of spin signals is of central interest for spintronics, where graphene stands as a relevant platform as its spin-orbit coupling (SOC) is tuned by proximity effects. Here, we propose an enhancement of spin-charge interconversion in graphene by controlling the intraparticle entanglement between the spin and pseudospin degrees of freedom. We demonstrate that, although the spin alone is not conserved in Rashba-Dirac systems, a combined spin-pseudospin operator is conserved. This conserved quantity represents the interconversion between pure spin and pseudospin textures to a spin-pseudospin entangled structure, where Kane-Mele SOC tunes this balance. By these means, we achieve spin-charge interconversion of 100\% efficiency via the Rashba-Edelstein effect. Quantum transport simulations in disordered micron-size systems demonstrate the robustness of this effect, and also reveal a disorder resilient spin Hall effect generated by the interplay between Rashba and Kane-Mele SOC. Our findings propose a platform for maximally efficient spin-charge interconversion, and establish spin-pseudospin correlations as a mechanism to tailor spintronic devices.
\end{abstract}

\maketitle

\section{Introduction}
Spin-charge interconversion (SCI) mechanisms lie at the core of spintronic technologies, generating current-driven spin signals which can then be used for information processing \cite{sinova_spin_2015, manchon_new_2015}. Traditionally, the development of efficient SCI relies on bulk heavy metals with a large spin-orbit coupling (SOC); however, van der Waals materials offer new possibilities as SOC can be tuned by proximity effects \cite{sierra_van_2021, yang_two-dimensional_2022, roche_spintronics_2024}. Graphene is particularly relevant in this context \cite{ahn_2d_2020, zhao_novel_2025}. 
Its long spin diffusion length makes it an ideal platform for spin transport and spin logic operations \cite{tombros_electronic_2007, zomer_long-distance_2012, wen_experimental_2016, khokhriakov_multifunctional_2022}; while SOC can be induced and tailored by proximity effects stacking graphene with transition metal dichalcogenides (TMDs) \cite{wang_strong_2015, wang_origin_2016, sun_determining_2023, rao_ballistic_2023, hoque_all-electrical_2021, tiwari_experimental_2022, sierra_room-temperature_2025} or topological insulators (TIs) \cite{khokhriakov_tailoring_2018, lee_proximity_2015, song_spin_2018, naimer_twist-angle_2023}. Efficient SCI via the Rashba-Edelstein effect (REE, also called inverse spin galvanic effect) and the spin Hall effect (SHE) has been shown in graphene-based heterostructures \cite{safeer_room-temperature_2019, ghiasi_charge-to-spin_2019, benitez_tunable_2020, khokhriakov_gate-tunable_2020}, even outperforming bulk heavy metals and non-van der Waals interfaces \cite{benitez_tunable_2020}; with the additional advantage of gate-tunability.

Most of graphene's unique features emerge from pseudospin related effects \cite{katsnelson_chiral_2006, morpurgo_intervalley_2006, mccann_weak-localization_2006, novoselov_two-dimensional_2005, zhang_experimental_2005}, where the pseudospin degree of freedom stemming from its bipartite sublattice structure generates the characteristic massless Dirac fermion behaviour. SOC in graphene is mediated by the pseudospin, and thus spin-pseudospin correlations play a central role in spin-related transport. Indeed, the weak localisation to weak antilocalisation transition in graphene is driven by both pseudospin and spin coherence effects \cite{garcia_spin_2017, sousa_weak_2022}; while the spin relaxation near the Dirac point is dominated by intra-particle spin-pseudospin entanglement \cite{tuan_pseudospin-driven_2014}. However, the role of spin-pseudospin correlations in SCI has not been fully understood. While graphene exhibits a unique vertical Rashba splitting enabling an efficient REE \cite{dyrdal_current-induced_2014, offidani_optimal_2017}, spin-pseudospin entanglement is most prominent in this regime \cite{offidani_optimal_2017, de-moraes_emergence_2020}. More recently, spin-pseudospin entanglement was proposed to drive spin-orbit torques in graphene beyond the electron gas regime generating unconventional spin-orbit torque fields \cite{sousa_skew-scattering-induced_2020, medina-duenas_emerging_2024}.

Here, we propose an enhanced SCI regime in graphene by controlling spin-pseudospin entanglement. We demonstrate the manipulation of spin-pseudospin entanglement by tuning different SOC fields; namely, Rashba and Kane-Mele (also called intrinsic) SOC. While spin angular momentum is not a conserved quantity due to Rashba SOC, we show that 
a combined spin-pseudospin operator is conserved, representing the interconversion between pure spin and pseudospin textures to a spin-pseudospin entangled structure, where Kane-Mele SOC tunes the spin-entanglement balance.
By these means, the REE is optimised up to maximal efficiency of 100\%, surpassing the previous theoretical maximum predicted in structures without Kane-Mele SOC \cite{offidani_optimal_2017}. We furthermore demonstrate the enhanced REE is highly robust by performing real-space simulations of micron-scale systems including disorder, maintaining a nearly maximal efficiency. Additionally, we demonstrate that Kane-Mele SOC reinstates a finite SHE in Rashba-Dirac systems even in presence of disorder, which is otherwise suppressed \cite{milletari_covariant_2017}.

\begin{figure*}[t]
    \includegraphics[width=\textwidth]{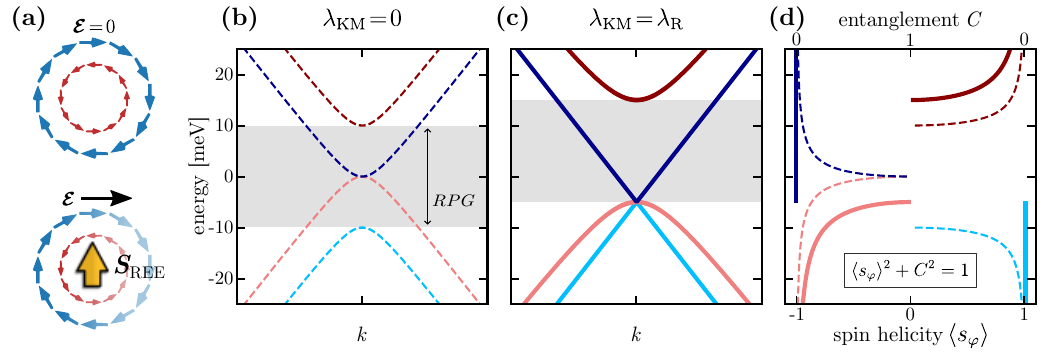}
    \caption{Depiction of the REE originating from helical spin textures in graphene. (a) Depiction of the REE effect, where the blue and red arrows represent the spin texture at their respective Fermi contours. Applying an electric field $\mathcal{E}$ generates a net spin density $\bm{S}_\text{REE} \propto \hat{\bm{z}} \times \bm{\mathcal{E}}$ (yellow arrow). (b, c) Band structure for (b) $\lambda_\text{KM} = 0$ and (c) $\lambda_\text{KM} = \lambda_\text{R} = 10 \,\text{meV}$. The Rashba pseudogap (RPG) is shaded in gray, while the light and dark red (blue) curves correspond to the $\chi=+$ ($\chi=-$) subspace. (d) Spin helicity (bottom axis) and spin-pseudospin entanglement (top axis) of the bands. Kane-Mele SOC reduces the effective mass in the blue cone, suppressing entanglement and enhancing the spin helicity throughout the pseudogap, leading to a larger REE.}
    \label{fig:fig1}
\end{figure*}

\section{Results and discussion}
In the REE, broken mirror symmetry parallel to the two-dimensional plane induces helical spin-momentum locking via Rashba SOC, where the spin texture winds about the Fermi contour orthogonal to the momentum. When applying a current, the non-equilibrium carrier redistribution generates a net spin density of the form $\bm{S}_\text{REE}^{} \propto \hat{\bm{z}} \times \bm{\mathcal{E}}$, with $\bm{\mathcal{E}}$ the driving electric field, as depicted in panel (a) of Fig. \ref{fig:fig1}. Vertical Rashba splitting in graphene generates favourable conditions for the REE as the current is carried through a single spin-helical band near the charge neutrality point \cite{offidani_optimal_2017}. However, because the pseudospin mediates spin-momentum locking, the spin texture is quenched by spin-pseudospin entanglement, representing the main limiting factor for the REE efficiency. Controlling spin-pseudospin entanglement thus emerges as the path towards optimizing SCI, as demonstrated in the following.

We consider a minimal graphene model endowed with Rashba SOC and Kane-Mele SOC. The low-energy Hamiltonian is
\begin{equation}
\begin{split}
    \mathcal{H}_{\bm{k}} &\, = \hbar v_\text{gr} \big( k_x \tau_z \sigma_x + k_y \sigma_y \big) - \frac{\lambda_\text{R}}{2} \big( \sigma_y s_x - \tau_z \sigma_x s_y \big) \\ &\hspace{1em} - \frac{\lambda_\text{KM}}{2} \tau_z \sigma_z s_z \text{ ,}
\end{split}
\label{eq:hamiltonian}
\end{equation}
with $v_\text{gr} \sim 10^6 \,\text{m} \,\text{s}^{-1}$ the velocity of massless electrons in bare graphene, and $\lambda_\text{R}^{}$ and $\lambda_\text{KM}^{}$ the Rashba and Kane-Mele SOC parameters respectively. The spin, pseudospin and valley degrees of freedom are respectively represented by the Pauli vectors $\bm{s}$, $\bm{\sigma}$ and $\bm{\tau}$. We will focus our analysis on the $\tau_z = +1$ subspace without loss of generality, as shown in the Supplementary Note S1. Graphene/TI heterostructures emerge as the most relevant platform for the proposed system, where both Rashba and Kane-Mele SOC are enhanced up to the meV scale \cite{khokhriakov_tailoring_2018, lee_proximity_2015, song_spin_2018, naimer_twist-angle_2023}; while in graphene/TMD bilayers Kane-Mele SOC is usually much smaller \cite{wang_strong_2015, wang_origin_2016, sun_determining_2023, rao_ballistic_2023, gmitra_graphene_2015}, but could be enhanced by adatom deposition \cite{weeks_engineering_2011, cresti_charge_2016}. The effects of other interactions which one may expect to encounter in such heterostructures, such as valley-Zeeman SOC or a staggered on-site potential, are analised in the Supplementary Note S2.

At large momentum the bare graphene Hamiltonian dominates, represented by the first term in Eq. \eqref{eq:hamiltonian}; and thus the bands are completely pseudospin polarised along the radial direction, $\langle \bm{\sigma} \rangle = \pm \hat{\bm{k}}$. Each subset of bands splits into clockwise and counter-clockwise spin helicity due to Rashba SOC, $\langle \bm{s} \rangle = \pm \hat{\bm{\varphi}} = \pm \hat{\bm{z}} \!\times\! \hat{\bm{k}}$, while Kane-Mele SOC does not influence the band structure in this regime. The resulting eigenstates, labelled by the radial pseudospin direction $\mu$ and spin helicity $\xi$ (with $\mu, \xi \!=\! \pm$), have the fully disentangled form $|\mu, \xi\rangle = |\mu \hat{\bm{k}}\rangle_{\sigma}^{} \otimes |\xi \hat{\bm{\varphi}}\rangle_{s}^{}$, where the first and second components correspond to the pseudospin and spin projected substates. Near the Dirac point, however, the SOC terms become of comparable strength to the bare graphene hamiltonian, resulting in the hybridisation of the asymptotic pure states and emergence of spin-pseudospin entanglement. 

We express the Hamiltonian in the asymptotic eigenbasis, revealing that the spectrum separates in two decoupled Dirac cones, one for each spin-pseudospin winding direction, labelled by $\chi \!=\! \,\text{sign}\, \langle \bm{\sigma} \times \bm{s} \rangle \cdot \hat{\bm{z}}$. The Hamiltonian projected to subspace $\chi$ reads
\begin{equation}
    \mathcal{H}_{\bm{k}, \chi} = \chi \frac{\lambda_\text{R}}{2} + \begin{pmatrix}
    - \hbar v_\text{gr} k & -\frac{1}{2}(\lambda_\text{KM} + \chi\lambda_\text{R}) \\
    -\frac{1}{2}(\lambda_\text{KM} + \chi\lambda_\text{R}) &  + \hbar v_\text{gr} k
    \end{pmatrix} \text{ ,}
    \label{eq:hamiltonian_chi}
\end{equation}
where the matrix elements correspond to the basis of pure states $|-, -\chi \rangle$ and $|+, +\chi \rangle$. The cones are gapped by an effective mass of magnitude $|\Delta_{\chi}| = |\lambda_\text{R} +\chi \lambda_\text{KM}|$, which promotes maximally entangled states of the form $\frac{1}{\sqrt{2}}(|-\!, -\chi \rangle \pm |+, +\chi \rangle)$ at $k=0$. The resulting bands are $E_{\chi, \pm} = \chi\lambda_\text{R}/2 \pm \varepsilon_\chi^{}$, with $\varepsilon_\chi^{} = [(\hbar v_\text{gr} k)^2 +  \Delta_\chi^2 / 4]^{1/2}_{}$, while the spin and pseudospin textures are $\langle \boldsymbol{s} \rangle_{\chi, \pm} = \pm\chi \rho_\chi \hat{\boldsymbol{\varphi}}$ and $\langle \boldsymbol{\sigma} \rangle_{\chi, \pm} = \pm \rho_{\chi} \hat{\boldsymbol{k}}$, with $\rho_\chi = \hbar v_\text{gr} k / \varepsilon_\chi$, where we have dropped the $\bm{k}$ index for simplicity (see Supplementary Note S1 for complete calculations). We note that Kane-Mele SOC, unlike other interactions such as a valley-Zeeman SOC, preserves the nature of the Rashba-Dirac system, marked by the spin-pseudospin winding of the bands.

The spin and pseudospin textures respectively remain fully helical and radial, but their magnitudes are quenched near the Dirac point. Because of SOC, the pure spin and pseudospin angular momenta are not conserved quantities, but these rather take the form of coupled spin-pseudospin operators. In particular we obtain the conserved quantity $\mathcal{Q} = \sigma_k s_\varphi$, with eigenvalues $\langle \mathcal{Q} \rangle_{\chi, \pm} = \chi$. The separable contribution to $\langle \mathcal{Q} \rangle$ is given by $\langle \mathcal{Q}\rangle_\text{sep} = \langle \sigma_k \rangle \langle s_\varphi \rangle$, and is determined by the pure spin and pseudospin textures. On the other hand, we find that the non-separable contribution, given by $\langle \mathcal{Q} \rangle - \langle \mathcal{Q} \rangle_\text{sep}$, is determined by the spin-pseudospin entanglement, as $\langle \sigma_k s_\varphi \rangle_{\chi, \pm} - \langle \sigma_k \rangle_{\chi, \pm} \langle s_\varphi \rangle_{\chi, \pm} = \chi C_{\chi,\pm}^2$, where $C$ is the concurrence, which quantifies entanglement from 0 to 1 from pure to maximally entangled states respectively \cite{wootters_entanglement_1998, de-moraes_emergence_2020} (see Supplementary Note S1 for explicit calculation). We conclude that the conserved quantity $\mathcal{Q} = \sigma_k s_\varphi$ represents the interconversion of angular momentum from the individual spin and pseudospin projections to spin-pseudospin entanglement. Finally, by expressing this relation as
\begin{subequations}
\begin{equation}
    \langle \bm{s}\rangle_{\chi, \pm}^2 + C_{\chi, \pm}^2 = 1 \text{ ,}
    \label{eq:sCconservation}
\end{equation}
\begin{equation}
    \langle \bm{\sigma}\rangle_{\chi, \pm}^2 + C_{\chi, \pm}^2 = 1 \text{ ,}
\end{equation}
\end{subequations}
we explicitly show that the entanglement is proportional to the quenching of the individual spin and pseudospin textures.

The spin texture throughout the Rashba pseudogap is enhanced by including Kane-Mele SOC, as shown in panels (b) and (c) of Fig.~\ref{fig:fig1}, for $\lambda_\text{KM} = 0$ and $\lambda_\text{KM} = \lambda_\text{R}$, respectively, while the light and dark red (blue) curves correspond to the $\chi\!=\!+$ ($\chi\!=\!-$) cone. The energy gap of the red $\chi\!=\!+$ (blue $\chi\!=\!-$) cone is crossed by the upper (lower) band of the opposite cone, forming the Rashba pseudogap, shaded in gray, where the Fermi level is intersected by a single spin helical band. Kane-Mele SOC opens the gap of the red $\chi\!=\!+$ cone, while that of the blue $\chi\!=\!-$ cone closes, reducing spin-pseudospin entanglement in the latter and enhancing the spin helicity throughout the pseudogap, as shown in panel (d). By these means, the spin texture can be tuned in the low-energy regime allowing to optimise the REE up to maximal efficiency, which we demonstrate below.

\begin{figure}[t]
    \centering
    \includegraphics[width=\linewidth]{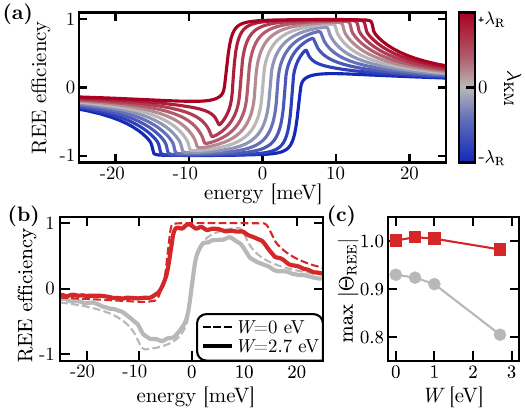}
    \caption{REE results. Panels (a) and (b) respectively show $\Theta_\text{REE}$ as a function of the Fermi energy in the clean limit and disordered system. The REE is optimised up to maximal efficiency $\Theta_\text{REE} = \pm 1$ for $\lambda_\text{KM} \!=\! \pm \lambda_\text{R}$ in the clean limit, remaining robust against disorder. Panel (c) shows the maximum $|\Theta_\text{REE}|$ as a function of the disorder strength $W$. [$\lambda_\text{R} = 10 \,\text{meV}$]}
    \label{fig:fig2}
\end{figure}

We quantify the REE efficiency adopting the definition proposed in Ref. \cite{offidani_optimal_2017}, which corresponds to the ratio between the transversal non-equilibrium spin density $S_y$, and the charge current $J_x$, yielding the efficiency $\Theta_\text{REE} = 2 e v S_y / (\hbar J_x)$. Here $v$ corresponds to the transport velocity required to adimensionalise the spin-to-current ratio.
Within a perturbative treatment of disorder, $v$ remounts to the Fermi velocity obtained from the band structure, $v_\text{F} = \hbar^{-1} \partial E /\partial k$, as initially considered in Ref. \cite{offidani_optimal_2017}. We extend this definition beyond the perturbative regime in order to further explore the robustness of SCI, as discussed in the Methods section.

We compute the non-equilibrium spin density at Fermi level $\varepsilon_\text{F}$ using the Kubo-Bastin formula \cite{bonbien_symmetrized_2020}, yielding
\begin{equation}
    S_y(\varepsilon_\text{F}) = \frac{\hbar \mathcal{E}}{\pi\Omega} \int \text{d}\varepsilon \,\mathit{f}(\varepsilon) \text{Im} \,\text{Tr} \Big[ \text{Im}G^+ \frac{\hbar}{2}s_y \frac{\partial G^+}{\partial \varepsilon} j_x \Big] \text{ ,}
    \label{eq:KB}
\end{equation}
with $\mathit{f}$ the Fermi-Dirac distribution (we take the zero temperature limit), $G^+ = \text{lim}_{\eta\rightarrow0^+} [ \varepsilon - H + \text{i}\eta]^{-1}$ the retarded Green's function, $\Omega$ the system size, $\mathcal{E}$ the driving electric field (taken along the $x$ axis), and $H$ and $\bm{j}$ the Hamiltonian and current operators respectively. The longitudinal current is computed identically, but replacing the spin operator $\frac{\hbar}{2} s_y$ by the current operator $j_x$.

In the clean limit the system is considered to be pristine, while a finite energy broadening $\eta$ is introduced in the Green's function, which generates a relaxation time $\tau_\eta = \hbar / 2 \eta$ representing a finite quasiparticle lifetime \cite{fan_linear_2021}. Both $S_y$ and $J_x$ are Fermi surface phenomena, proportional to $\tau_\eta$, and therefore the REE efficiency is independent of the chosen broadening.
We obtain for the non-equilibrium expectation values
\begin{subequations}
\label{eq:noneqSandJ}
\begin{equation}
    S_y (\varepsilon_\text{F}) = -\frac{e \mathcal{E} \tau_\eta}{4\pi\hbar} \sum_{k_\text{F}} \, k \, \frac{\hbar}{2}\langle s_\varphi \rangle \, \text{sign}\, \langle \sigma_k \rangle \text{ ,}
\end{equation}
\begin{equation}
    J_x (\varepsilon_\text{F}) = \frac{e^2 v_\text{gr} \mathcal{E} \tau_\eta}{4\pi \hbar} \sum_{k_\text{F}} k \, |\langle \sigma_k \rangle| \text{ ,}
\end{equation}
\end{subequations}
where the sum is performed over all Fermi momenta. Within the pseudogap there is a single Fermi momenta and $\Theta_\text{REE}$ is uniquely determined by the spin helicity of the gap-crossing band, i.e. $\Theta_\text{REE} = \langle s_\varphi \rangle \,\text{sign}\langle \sigma_k \rangle |_{k_\text{F}}^{}$; while outside the pseudogap bands of opposite helicities counteract. Thus, the REE is maximised at the pseudogap edges, yielding
\begin{equation}
    \text{max}\, \Theta_\text{REE} = \frac{2 \sqrt{2 \lambda_\text{R} (\lambda_\text{R} + \lambda_\text{KM})}}{3\lambda_\text{R} +  \lambda_\text{KM}} \text{ ,}
\end{equation}
where we have considered $\lambda_\text{R}, \lambda_\text{KM} \geq 0$. $\Theta_\text{REE} (\varepsilon_\text{F})$ is shown in Fig. \ref{fig:fig2}-(a) for different values of $\lambda_\text{KM}$, demonstrating the enhancement of the REE due to Kane-Mele SOC. Not only does the maximum value of $|\Theta_\text{REE}|$ increase with $\lambda_\text{KM}$, but it also peaks throughout a wider spectral range. Remarkably, the REE is optimised up to maximal efficiency of $\Theta_\text{REE} = \pm 1$ for $\lambda_\text{KM} = \pm \lambda_\text{R}$ due to the complete suppression of spin-pseudospin entanglement throughout the Rashba pseudogap, surpassing the previous theoretical maximum \cite{offidani_optimal_2017}.

Beyond the clean limit, we demonstrate the robustness of the enhanced REE by performing real-space simulations treating impurity scattering in a non-perturbative fashion. 
Because we intend to discern transport phenomena within the Rashba pseudogap, we require a spectral resolution on the meV scale, where a large system size is necessary to avoid spurious finite-size effects. For this, we compute Eq. \eqref{eq:KB} by the Kernel Polynomial method, developing an energy-filtering technique \cite{weisse_kernel_2006, fan_linear_2021, gimenez-de-castro_fast_2024, gimenez-de-castro_efficient_2023}, which allows us to simulate a graphene sheet of $\sim 28 \, \mu\text{m}^2$ ($\sim10^9$ atoms) with a numerical Jackson spectral broadening of $ \eta \sim 1 \,\text{meV}$ (see Methods section below, and complete results in Supplementary Note S3). Anderson disorder is included as random on-site potentials in the range $[-W/2, W/2]$. The relaxation time is now given by $\tau^{-1} = \tau_\eta^{-1} + \tau_p^{-1}$, where $\tau_\eta$ is methodology-induced by $\eta$, while the disorder-induced relaxation is represented by $\tau_p$.

Our results, shown in Fig. \ref{fig:fig2}-(b, c), reveal a remarkable robustness of the REE. Indeed, both $S_y$ and $J_x$ remain mainly unaffected by disorder within the pseudogap, while they are suppressed outside, revealing a stronger robustness to disorder in the former regime, granted by a larger $\tau_p$. This behaviour occurs for all values of the SOC parameters, where the SCI enhancement due to Kane-Mele SOC persists. The largest efficiency losses occur at the pseudogap edges, where disorder-induced scattering enables transitions between bands of opposite helicities, thus breaking the Rashba pseudogap regime. For $\lambda_\text{KM} \!=\! 0$ we observe that the $\Theta_\text{REE}$ peak, which in the clean limit occurs at the RPG edges, is reduced by the disorder-induced smearing of the RPG. On the other hand, for $\lambda_\text{KM}=\lambda_\text{R}$ the plateau-shaped maximum grants further robustness to the REE.

We now show that SCI via the SHE is also enhanced by modifying spin-pseudospin entanglement. While in the Kane-Mele dominated system the SHE is topological and robust to disorder \cite{kane_quantum_2005, sheng_nondissipative_2005}, in a pure Rashba-Dirac system the intrinsic contribution to the SHE is countered by disorder \cite{milletari_covariant_2017, daSilva_spin_2022}. Therefore, we focus on the $\lambda_\text{KM} < \lambda_\text{R}$ regime. Theoretical calculations in the clean limit have shown the emergence of a peak in the SHE within the RPG due to Kane-Mele SOC \cite{dyrdal_spin_2009}; however, its response against disorder remains unexplored. To address this issue we compute the spin Hall current in real-space disordered systems using Eq. \eqref{eq:KB}, replacing $\hbar s_y / 2$ for the spin current operator $j_y^z = \hbar / 4 \, \{ v_y, s_z \}$, where $\bm{v}$ corresponds to the velocity operator. 

\begin{figure}[t]
    \centering
    \includegraphics[width=\linewidth]{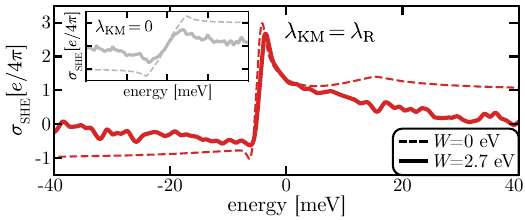}
    \caption{SHE conductivity, $\sigma_\text{SHE}$, for $\lambda_\text{KM}=\lambda_\text{R}$ (main panel) and $\lambda_\text{KM} = 0$ (inset), without disorder and with disorder of $W=2.7 \,\text{eV}$, respectively shown in dashed and solid lines. [$\lambda_\text{R} = 10 \,\text{meV}$]}
    \label{fig:fig3}
\end{figure}

Our results for the spin Hall conductivity $\sigma_\text{SHE}$ are shown in Fig. \ref{fig:fig3}, with and without Kane-Mele SOC in the main panel and inset, respectively. For the pure Rashba system our real-space simulations capture the disorder-induced suppression of the intrinsic SHE, as shown in the inset of Fig. \ref{fig:fig3}, in accordance with theoretical predictions based on covariant conservation laws and diagrammatic analysis \cite{milletari_covariant_2017}. 
When including Kane-Mele SOC, calculations in the clean limit show that outside the RPG the SHE remains similar to that of the pure Rashba system, while a prominent $\sigma_\text{SHE}^{}$ peak emerges from the band touching at charge neutrality. The inter-band transition originating this peak occurs between Dirac cones of opposite spin-pseudospin winding $\chi$, which we find is proportional to the difference in entanglement between both Dirac cones (see Supplementary note S1.). The $\sigma_\text{SHE}$ peak is highly robust to disorder, as shown in Fig. \ref{fig:fig3}; while, similar to the pure Rashba system, away from the charge neutrality point the SHE is suppressed. By these means, the interplay between Rashba and Kane-Mele SOC reinstates a finite SHE in a Rashba-Dirac system, even in the Rashba dominated regime where there is no topological quantum spin Hall effect gap. 
This result resembles that obtained for valley-Zeeman SOC, where the interplay with the Rashba field yields a finite SHE \cite{milletari_covariant_2017}. However, we note that in the latter case intervalley scattering strongly suppresses the SHE \cite{garcia_spin_2017}, while in the present system both intra and intervalley scattering act on the same level.

Altogether, our results unveil the role of spin-pseudospin correlations for spin-charge interconversion mechanisms in graphene.
In particular, we have shown that spin-pseudospin entanglement can serve as a robust tool both to optimise spin-momentum locking in order to enhance the Rashba-Edelstein effect up to maximal efficiency; as well as to tune inter-band transitions generating a finite spin Hall effect. While we have here provided a proof of concept from a minimal graphene model, further research along this path may address these phenomena in other Dirac materials, as well as discern the effects of spin-pseudospin correlations for spin-orbit torque in magnetic devices.

\section{Methods}

\subsection{Quantum transport: framework}
Under a driving electric field, the macroscopic non-equilibrium density of some observable $\mathcal{O}$, is given by Eq. \eqref{eq:KB}, replacing the spin operator $\frac{\hbar}{2}s_y$ for $\mathcal{O}$. This expression may be further separated in Fermi surface and Fermi sea contributions \cite{bonbien_symmetrized_2020}
\begin{subequations}
\begin{equation}
    \mathcal{O}_\text{surf}(\varepsilon_\text{F}) = \frac{\hbar}{\pi \Omega} \int \text{d}\varepsilon \Big( -\frac{\partial f}{\partial \varepsilon} \Big) \,\text{Tr}\, \Big[ \text{Im}\, G^+ \, \mathcal{O} \, \,\text{Im}\,G^+ \, \big( \bm{\mathcal{E}} \cdot \bm{j} \big)\Big] \text{ ,}
\end{equation}
\begin{equation}
    \mathcal{O}_\text{sea}(\varepsilon_\text{F}) = \frac{2\hbar}{\pi \Omega} \int \text{d}\varepsilon f \,\,\text{Im} \,\text{Tr}\, \Big[ \text{Im}\, G^+ \, \mathcal{O} \, \frac{\partial \,\text{Re}\, G^+}{\partial \varepsilon} \, \big( \bm{\mathcal{E}} \cdot \bm{j} \big)\Big] \text{ .}
\end{equation}
\end{subequations}

In the clean limit the system is considered to be periodic, while a finite yet small broadening $\eta$ is introduced in the Green's function. The trace is taken in reciprocal space as $\Omega^{-1}\,\text{Tr} \rightarrow \sum_n \int \text{d}^2\bm{k} / (2\pi)^2$, where the sum is performed over all eigenstates. For the Fermi surface contribution we use $\sum_{n, m} \text{Im}\, G^+(E_n) \,\text{Im}\, G^+(E_m) = \sum_n (\text{Im}\, G^+(E_n))^2$, and $\int \text{d}\varepsilon\, \eta^2 / (\varepsilon^2 + \eta^2)^{-2} = \pi/2\eta$ to obtain
\begin{equation}
    \mathcal{O}_\text{surf} (\varepsilon_\text{F}) =  \tau_\eta \sum_n \int \frac{\text{d}^2\bm{k}}{(2\pi)^2} \delta(E_{n, \bm{k}} - \varepsilon_\text{F}) \langle \mathcal{O} \rangle_{n, \bm{k}} \langle \bm{\mathcal{E}} \cdot \bm{j} \rangle_{n, \bm{k}}
    \text{ ,}
    \label{eq:SM_KBsurf}
\end{equation}
with $\tau_\eta = \hbar / 2\eta$ the $\eta$-induced relaxation time. The Fermi sea contribution on the other hand is finite in the $\eta \rightarrow 0^+$ limit, given by
\begin{equation}
\begin{split}
    \mathcal{O}_\text{sea} =&\, 2\hbar \sum_{n, m} \int \frac{\text{d}^2\bm{k}}{(2\pi)^2} f(E_n) \\& \frac{\text{Im}\, \langle E_{n, \bm{k}} | \mathcal{O} | E_{m, \bm{k}} \rangle \langle E_{m, \bm{k}} | \bm{\mathcal{E}} \cdot \bm{j} | E_{n, \bm{k}} \rangle}{(E_{n, \bm{k}} - E_{m, \bm{k}})^2 + \eta^2} \text{ .}
    \label{eq:SM_KBsea}
\end{split}
\end{equation}
The REE spin density and longitudinal current obtained in Eq. \eqref{eq:noneqSandJ} are respectively calculated from Eq. \eqref{eq:SM_KBsurf} by replacing $\mathcal{O} \rightarrow \frac{\hbar}{2} s_y$ and $\mathcal{O} \rightarrow j_x = -ev_\text{gr} \sigma_x$, where we have assumed $\bm{\mathcal{E}} = \mathcal{E} \hat{\bm{x}}$ exploiting the rotational symmetry of the system. The intrinsic spin Hall current, on the other hand, is a Fermi sea phenomena, obtained by replacing $\mathcal{O} \rightarrow j_y^z = \hbar/4 \{ v_y, s_z \}$ in Eq. \eqref{eq:SM_KBsea}.

\subsection{Quantum transport: numerics}
For real-space calculations in disordered systems we use the tight-binding graphene Hamiltonian
\begin{equation}
\begin{split}
    H =&\, \sum_{\langle i, j \rangle} -\gamma_0 c_i^\dagger c_j + \frac{\text{i} \lambda_\text{R}}{3}c_i^\dagger \bm{s} \cdot \Big(\hat{\bm{z}} \times \frac{\bm{r}_{i,j}}{a_\text{C-C}} \Big) c_j + \\&\, \sum_{\langle \langle i , j \rangle \rangle} \frac{\text{i} \lambda_\text{KM}}{6\sqrt{3}} c_i^\dagger s_z c_j \, \alpha_i \,\text{cos} \Big( 3 \,\text{arccos}\, \frac{ \bm{r}_{ \text{A} \rightarrow \text{B}} \cdot \bm{r}_{i,j}}{\sqrt{3} a_\text{C-C}^2} \Big) + \\&\,\sum_i w_i c_i^\dagger c_i
    \text{ ,}
\end{split}
\end{equation}
where $c_i^\dagger$ is the electron spinor creation operator at position $\bm{r}_i$, $\bm{r}_{i,j} = \bm{r}_i - \bm{r}_j$, and $\bm{r}_{\text{A} \rightarrow \text{B}}$ is a vector pointing from a site of sublattice A to its nearest neighbour of sublattice B, with $a_\text{C-C} = 0.142 \,\text{nm}$ the nearest neighbour distance, and $\alpha_i=\pm$ for $c_i^\dagger$ of sublattice A(B). Here $\gamma_0 = 2.7 \,\text{eV}$ is the nearest neighbour hopping, $\lambda_\text{R}$ and $\lambda_\text{KM}$ are the Rashba and Kane-Mele SOC parameters respectively, and $w_i$ is a random number from the uniform distribution $[-W/2, W/2]$ with $W$ the Anderson disorder parameter.

Transport calculations are performed using the full Kubo-Bastin equation, given by Eq. \eqref{eq:KB}, solved by the kernel Polynomial method (KPM) \cite{weisse_kernel_2006, garcia_real-space_2015}. The Green's function is expanded in Chebyshev polynomials using the Jackson kernel to damp the Gibbs oscillations, where the $\eta \rightarrow 0^+$ limit is taken; however, a spectral broadening is generated by the expansion. The resulting broadening parameter is $\eta_\text{KPM} \approx \pi \Delta E/M$ at the band centre, with $\Delta E$ the bandwidth and $M$ the order of the Chebyshev expansion \cite{weisse_kernel_2006}. To compute transport properties in the meV scale near the Dirac point we use an energy filtering technique allowing us to reach $\eta_\text{KPM} \sim 1 \,\text{meV}$ with $M=800$, with a system size as large as $10^9$ atoms \cite{medina-duenas_emerging_2024}. Similar to the clean limit, the quasi-particle lifetime is given by $\tau_\eta(\varepsilon_\text{F}) = \text{lim}_{\eta\rightarrow 0^+} \frac{\hbar}{\pi} \int \text{d}\varepsilon\,  \eta^2 / ((\varepsilon_\text{F} - \varepsilon)^2 + \eta^2)^2$, where the Lorentzian function must be expanded in Chebyshev polynomials.  We obtain
\begin{equation}
\begin{split}
    \tau_\eta(\varepsilon_\text{F}) =&\, \frac{\hbar}{\pi \Delta E} \Bigg( \sum_{m_1, m_2 \text{ even}} + \sum_{m_1, m_2 \text{ odd}} \Bigg) g_{m_1} g_{m_2} \\& T_{m_1}(\tilde{\varepsilon}_\text{F}) T_{m_2}(\tilde{\varepsilon}_\text{F}) \frac{(2 - \delta_{0, m_1}) (2 - \delta_{0, m_2})}{1 - \tilde{\varepsilon}_\text{F}^2} 
    \\&\, \Bigg( \frac{1}{1 - (m_1 + m_2)^2} + \frac{1}{1 - (m_1 - m_2)^2} \Bigg)
    \\\approx&\, 1.84 \frac{\hbar}{2 \eta(\varepsilon_\text{F})}
    \text{ ,}
\end{split}
\end{equation}
with $g_m$ the Jackson kernel coefficients, $T_m$ the Chebyshev polynomials and $\tilde{\varepsilon}_\text{F}$ the normalised $\varepsilon_\text{F}$ within the values $[-1, 1]$ by the bandwidth, and where the second equality was obtained by numerical calculation.

\subsection{Transport velocity}
In order to compute the REE efficiency in real-space disordered systems, we must construct a definition of the transport velocity which does not remount to band structure properties, but rather relies on macroscopic quantities which can be computed by the KPM simulations. Let's begin from the clean limit. The conductivity is given by Eq. (5b) of the main text, while the density of states is $\rho(\varepsilon_\text{F}) = (\pi \Omega)^{-1} \text{Tr}\, \text{Im}\, G^+(\varepsilon_\text{F}) = (2\pi)^{-1} \sum_{k_\text{F}} k (\hbar v_\text{gr} \langle \sigma_k \rangle)^{-1}$. The band structure Fermi velocity $v_\text{F} = \hbar^{-1} \partial E / \partial k = v_\text{gr} \langle \sigma_k \rangle$ can thus be obtained as $v(\varepsilon_\text{F}) = \sqrt{2\sigma(\varepsilon_\text{F}) / e^2 \rho(\varepsilon_\text{F}) \tau_\eta(\varepsilon_\text{F})}$ within the RPG. We now offer an interpretation of this quantity in terms of the carrier wave packet evolution. We may use the Einstein relation to write the conductivity as $\sigma(\varepsilon_\text{F}) = e^2 \rho(\varepsilon_\text{F}) D$ , with $D(\varepsilon_\text{F}) = \lim_{t\rightarrow \infty} \Delta X^2(\varepsilon_\text{F}, t) / 2t$ the diffusion coefficient, where $\Delta X^2(\varepsilon_\text{F}, t)$ is the mean square displacement of the wave packet, and $t$ its evolution time. The largest time scale accessible within the KPM simulations corresponds to $\tau_\eta$, which can be interpreted as the carrier quasiparticle coherence time \cite{fan_linear_2021}. We thus take $D(\varepsilon_\text{F}) = \Delta X^2(\varepsilon_\text{F}, \tau_\eta) / 2\tau_\eta$. With this, the transport velocity reads $v(\varepsilon_\text{F}) = \sqrt{\Delta X^2(\varepsilon_\text{F}, \tau_\eta)} / \tau_\eta$, which we interpret as the mean velocity of the carriers throughout their evolution.

\section{Code availability}
The computer codes used to produce the numerical results are available
from the corresponding authors upon reasonable request.

\section{acknowledgments}
We thank Aires Ferreira for fruitful discussions on the state of the art of SCI.
The authors acknowledge funding from: project I+D+i PID2022-138283NB-I00 funded by MICIU/AEI/10.13039/501100011033/ and “FEDER Una manera de hacer Europa”; FLAG-ERA project MNEMOSYN (PCI2021-122035-2A) funded by MICIU/AEI /10.13039/501100011033 and the European Union NextGenerationEU/PRTR; ERC project AI4SPIN funded by Horizon Europe-European Research Council Executive Agency under grant agreement No 101078370-AI4SPIN; MICIU with European funds‐NextGenerationEU (PRTR‐C17.I1); and 2021 SGR 00997 funded by Generalitat de Catalunya.
J.M.D. acknowledges support from MICIU grant FPI PRE2021-097031. 
ICN2 is funded by the CERCA Programme/Generalitat de Catalunya and supported by the Severo Ochoa Centres of Excellence programme, Grant CEX2021-001214-S, funded by MCIN/AEI/10.13039.501100011033.

\section{Author contributions}
J.~M.~D. designed the research, obtained main theoretical results, and wrote the manuscript. S.~G.~de~C. developed and ran the computer code to obtain numerical results. J.~H.~G. and S.~R. co-supervised the project.

\section{Competing interests}
The authors declare no competing interests.

\end{document}


\title{Supplementary information for ``Optimal spin-charge interconversion in graphene through spin-pseudospin entanglement control''} 

\author{Joaqu{\'i}n Medina Due{\~n}as}
\email[Corresponding Author ]{joaquin.medina@icn2.cat}
\affiliation{ICN2 --- Catalan Institute of Nanoscience and Nanotechnology, CSIC and BIST, Campus UAB, Bellaterra 08193, Barcelona, Spain}
\affiliation{Department of Physics, Universitat Aut{\`o}noma de Barcelona (UAB), Campus UAB, Bellaterra, 08193 Barcelona, Spain}

\author{Santiago Gim{\'e}nez de Castro}
\affiliation{ICN2 --- Catalan Institute of Nanoscience and Nanotechnology, CSIC and BIST, Campus UAB, Bellaterra 08193, Barcelona, Spain}

\author{Jos{\'e} H. Garc{\'i}a}
\email[Corresponding Author ]{josehugo.garcia@icn2.cat}
\affiliation{ICN2 --- Catalan Institute of Nanoscience and Nanotechnology, CSIC and BIST, Campus UAB, Bellaterra 08193, Barcelona, Spain}

\author{Stephan Roche}
\affiliation{ICN2 --- Catalan Institute of Nanoscience and Nanotechnology, CSIC and BIST,
Campus UAB, Bellaterra 08193, Barcelona, Spain}
\affiliation{ICREA --- Instituci\'o Catalana de Recerca i Estudis Avan\c{c}ats, 08010 Barcelona, Spain}

\maketitle

\onecolumngrid

\section{Electronic structure: Analytical developments}
We begin from the Hamiltonian in Eq. (1) of the main text. We transform the hamiltonian as $\mathcal{H}_{\bm{k}} \rightarrow U^\dagger \mathcal{H}_{\bm{k}} U$, with $U = \tau_+^{} + \tau_-^{} \sigma_y$ and $\tau_\pm = (\tau_0 \pm \tau_z)/2$. Now the Hamiltonian at both valleys presents the same form, and we may focus our analysis only on the $\tau_z=+1$ subspace. We express the Hamiltonian in the basis $\{ |-,+ \rangle \,,\, |+,- \rangle \,,\, |-,- \rangle \,,\, |+,+ \rangle \}$, where $|\mu,\xi\rangle$ corresponds to a pure state with pseudospin and spin respectively aligned along $\mu\hat{\bm{k}}$ and $\xi \hat{\bm{\varphi}}=\xi \hat{\bm{z}} \times \hat{\bm{k}}$, obtaining
\begin{equation}
\begin{split}
    \mathcal{H}_{\bm{k}} =&\, \begin{pmatrix}
        -\frac{1}{2}\lambda_\text{R} - \hbar v_\text{gr} k & -\frac{1}{2} (\lambda_\text{KM} - \lambda_\text{R}) & 0 & 0 \\
        -\frac{1}{2} (\lambda_\text{KM} - \lambda_\text{R}) & -\frac{1}{2}\lambda_\text{R} + \hbar v_\text{gr} k & 0 & 0 \\
        0 & 0 & \frac{1}{2}\lambda_\text{R} - \hbar v_\text{gr} k & -\frac{1}{2} (\lambda_\text{KM} + \lambda_\text{R}) \\
        0 & 0 & -\frac{1}{2} (\lambda_\text{KM} +  \lambda_\text{R}) & +\frac{1}{2}\lambda_\text{R} + \hbar v_\text{gr} k
    \end{pmatrix} \\
    =&\, \bigoplus_{\chi=\mp} \mathcal{H}_{\bm{k}, \chi}^{} \text{ ,} 
\end{split}
\end{equation}
with $\mathcal{H}_{\bm{k}, \chi}$ given by Eq. (2) of the main text. Comparing with a simple gapped graphene model $\mathcal{H}_\text{gr} = \hbar v_\text{gr} \bm{k} \cdot \bm{\sigma} + \sigma_z \Delta / 2$, we note that $\mathcal{H}_{\bm{k},\chi}$ corresponds to a Dirac cone with an effective mass of $\Delta_\chi = -(\lambda_\text{KM} + \chi \lambda_\text{R})$, with the additional feature of helical spin-pseudospin locking. The eigenvalues and eigenvectors of $\mathcal{H}_{\bm{k}, \chi}$ are
\begin{subequations}
\begin{equation}
    E_{\chi, \pm} = \chi\frac{\lambda_\text{R}}{2} \pm \varepsilon_\chi \text{ ,} 
\end{equation}
\begin{equation}
    | E_{\chi, \pm} \rangle = \frac{1}{\sqrt{2 \varepsilon_{\chi}}} \begin{pmatrix}
        -\text{sign}(\Delta_\chi) \sqrt{\varepsilon_\chi \mp \hbar v_\text{gr} k} \\
        \pm \sqrt{\varepsilon_\chi \pm \hbar v_\text{gr} k}
    \end{pmatrix} \text{ ,}
    \label{eq:SM_evecs}
\end{equation}
\end{subequations}
with $\varepsilon_\chi = [(\hbar v_\text{gr} k)^2 + \Delta_\chi^2 / 4]^{1/2}$, and where $|E_{\chi, \pm}\rangle$ is represented in basis $\{ |-,-\chi \rangle \,,\, | +,+\chi \rangle \}$, and we have omitted the $\bm{k}$ subindex for simplicity. The spin and pseudospin expectation values for the eigenstates are $\langle \bm{s} \rangle_{\chi, \pm} = \pm \chi \rho_\chi \hat{\bm{\varphi}}$ and $\langle \bm{\sigma} \rangle_{\chi, \pm} = \pm \rho_\chi \hat{\bm{k}}$, with $\rho_{\chi} = \hbar v_\text{gr} k / \varepsilon_\chi$.

Even though the spin and pseudospin textures remain fully helical and radial, respectively, we find that these do not correspond to conserved quantities. Indeed, calculating the commutators we find
\begin{subequations}
\begin{equation}
    [\mathcal{H} \,,\, s_\varphi] = -\text{i}\lambda_\text{R} \sigma_\varphi s_z + \text{i}\lambda_\text{KM} \sigma_z s_k \text{ ,}
\end{equation}
\begin{equation}
    [\mathcal{H} \,,\, \sigma_k] = -\text{i}\lambda_\text{KM} \sigma_\varphi s_z + \text{i}\lambda_\text{R} \sigma_z s_k \text{ .}
\end{equation}
\end{subequations}
Additionally, we have $[\mathcal{H}, \sigma_z s_z] = -2\text{i} \hbar v_\text{gr} k \sigma_\varphi s_z$ and $[\mathcal{H}, \sigma_\varphi s_k] = 2\text{i} \hbar v_\text{gr} k \sigma_z s_k$, From this we obtain the conserved quantities, \begin{subequations}
\begin{equation}
    \mathcal{Q}_1 = s_\varphi - \frac{1}{2\hbar v_\text{gr} k}(\lambda_\text{R} \sigma_z s_z + \lambda_\text{KM} \sigma_\varphi s_k) \text{ ,}
\end{equation}
\begin{equation}
    \mathcal{Q}_2 = \sigma_k - \frac{1}{2\hbar v_\text{gr} k}(\lambda_\text{KM} \sigma_z s_z + \lambda_\text{R} \sigma_\varphi s_k) \text{ ,}
\end{equation}
\end{subequations}
along with $\mathcal{Q}_0 = \sigma_k s_\varphi$ already analysed in the main text.

We quantify entanglement using the 
concurrence, which is computed as follows. For a pair of coupled two-level systems, the concurrence is defined as $C_\psi = |\langle \psi | \sigma_y^\text{A} \sigma_y^\text{B} | \psi^*\rangle|$, where the quantisation axis within both subsystems is $\hat{\bm{z}}$, $\sigma_y^{\text{A(B)}}$ is the Pauli matrix $\sigma_y$ in subsystem A(B), and $|\psi^*\rangle$ is the complex conjugated wavefunction. We employ the same definition, but quantising the spin and pseudospin subspaces along $\hat{\bm{\varphi}}$ and $\hat{\bm{k}}$ respectively, where additionally we note that the eigenstates only present real coefficients as seen in Eq. \eqref{eq:SM_evecs}. The concurrence for the eigenstates is thus $C_{\chi,\pm} = |\langle \sigma_\varphi s_k \rangle_{\chi,\pm}| = |\Delta_\chi| / 2 \varepsilon_\chi$. We note that $\langle \sigma_\varphi \rangle_{\chi, \pm} = \langle s_k \rangle_{\chi, \pm} = 0$, and thus $\langle \sigma_\varphi s_k \rangle_{\chi, \pm}$ reflects the non-separability of the wavefunction. Additionally, expressing $\sigma_z s_z$ as a rotation of $\sigma_\varphi s_k$, we obtain $\langle \sigma_z s_z \rangle_{\chi,\pm} = \chi \langle \sigma_\varphi s_k \rangle_{\chi, \pm}$. The conserved quantities can now be written as $\langle \mathcal{Q}_1 \rangle_{\chi, \pm} = \langle s_\varphi \rangle_{\chi, \pm} - \Delta_\chi \langle \sigma_\varphi s_k \rangle_{\chi, \pm} / 2\hbar v_\text{gr} k$, and $\langle \mathcal{Q}_2 \rangle_{\chi, \pm} = \langle \sigma_k \rangle_{\chi, \pm} - \chi \Delta_\chi \langle \sigma_\varphi s_k \rangle_{\chi, \pm} / 2\hbar v_\text{gr} k$. By these means, $\mathcal{Q}_1$ and $\mathcal{Q}_2$ are also interpreted as the interconversion of angular momentum from the individual spin and pseudospin projections to spin-pseudospin entanglement.

\section{Electronic structure: Other Hamiltonian terms}

In real graphene-based heterostructures other proximity-induced interactions may also be relevant. 
We analyse the case of a staggered onsite potential and valley Zeeman SOC, which are relevant in graphene/TMD interfaces due to the breaking of sublattice symmetry \cite{gmitra_graphene_2015}; as well as the case of a Kekulé-O distortion which is relevant in graphene/topological insulator interfaces \cite{song_spin_2018}. These interactions are represented in the Hamiltonian by respectively including the terms $\mathcal{H}_\text{st}$, $\mathcal{H}_\text{VZ}$ and $\mathcal{H}_\text{Kek}$, given by
\begin{subequations}
\begin{equation}
    \mathcal{H}_\text{st} = \frac{1}{2} M \sigma_z \rightarrow \frac{1}{2} M \tau_z \sigma_z \text{} ,
\end{equation}
\begin{equation}
    \mathcal{H}_\text{VZ} = -\frac{1}{2}\lambda_\text{VZ} \tau_z s_z \rightarrow -\frac{1}{2}\lambda_\text{VZ} \tau_z s_z \text{ ,}
\end{equation}
\begin{equation}
    \mathcal{H}_\text{Kek} = \Delta \tau_y \sigma_x \rightarrow \frac{1}{2}\Delta \tau_x \sigma_z \text{ ,}
\end{equation}
\end{subequations}
where the right arrow indicates the unitary transformation $U$. We now consider the effects of these interactions for charge-to-spin conversion, calculated in the clean limit.

\begin{figure}[b]
    \centering
    \includegraphics[width=\linewidth]{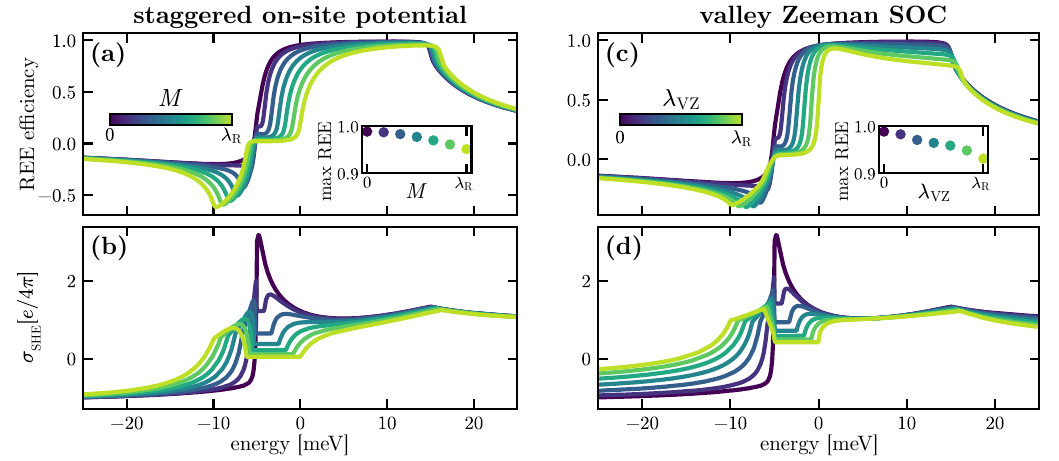}
    \caption{REE efficiency and spin Hall conductivity ($\sigma_{xy}^z$), including a staggered on-site potential $\mathcal{H}_\text{st}$ and valley Zeeman SOC $\mathcal{H}_\text{VZ}$. [$\lambda_\text{KM} = \lambda_\text{R} = 10 \,\text{meV}$]}
    \label{fig:SM_CSC}
\end{figure}

The staggered potential introduces a hybridisation between the states $|- ,\xi\rangle$ and $|+, \xi\rangle$ (for $\xi=\pm$), of opposite sign in each valley. At $k=0$ a gap of magnitude $|M|$ opens between bands $E_{-,+}$ and $E_{++}$ which brings both the pseudospin and spin textures out of the plane; while for larger momentum the interaction is suppressed due to the increasing spectral distance between the bands, and the textures approach the $M=0$ case. Fig. \ref{fig:SM_CSC}-(a, b) shows charge to spin conversion for different $M$ values, with $\lambda_\text{KM} = \lambda_\text{R}$. We observe that the REE is more strongly suppressed at the low-energy end of the pseudogap, where the Fermi momentum is near $k=0$, while at the high-energy end it is closer to the $M=0$ case because the Fermi momentum is larger. The peak REE value is 0.952 for $M = \lambda_\text{R}$, showing that the REE remains largely robust against a staggered on-site potential. In the SHE, shown in panel (b), the gap opening regularises the diverging peak at $k=0$, where a finite SHE appears within the gap and tends to zero for increasing $M$, as the gap is dominated by the topologically trivial staggered potential.

Valley Zeeman SOC, on the other hand, introduces a hybridisation between states of the same pseudospin and opposite spin polarisations, $|\mu,+\rangle$ and $|\mu,-\rangle$ (for $\mu=\pm$). The spectral distance between these states remains fairly constant throughout the spectrum, and thus the interaction is not suppressed at large momentum. The eigenstates acquire a finite out-of-plane spin texture component which suppresses Rashba-related effects, as can be seen by the reduction of the asymptotic SHE value, and the REE at the high-energy end of the pseudogap, shown in Fig. \ref{fig:SM_CSC}-(d) and (c), respectively. At $k=0$ the gap opening regularises the diverging peak in the SHE, similar to the case of the staggered potential. The REE, on the contrary, reaches its peak value near the low-energy end of the pseudogap, where the out-of-plane component of the spin texture is zero. For $\lambda_\text{VZ} = \lambda_\text{R}$ the peak REE value is of 0.932, also showing robustness against valley Zeeman SOC.

For the Kekulé-O distortion we note that under a $\pi$ rotation about $\tau_y$, $\mathcal{H}_\text{Kek}$ takes the same form as $\mathcal{H}_\text{st}$, while the rest of the Hamiltonian terms remain unaffected. The Kekulé-O distortion thus presents the same effects as the staggered on-site potential, where the eigenstates are valley-polarised along $\tau_x$ rather than $\tau_z$.

\section{Transport calculations: complete data}

\begin{figure}
    \centering
    \includegraphics[width=\linewidth]{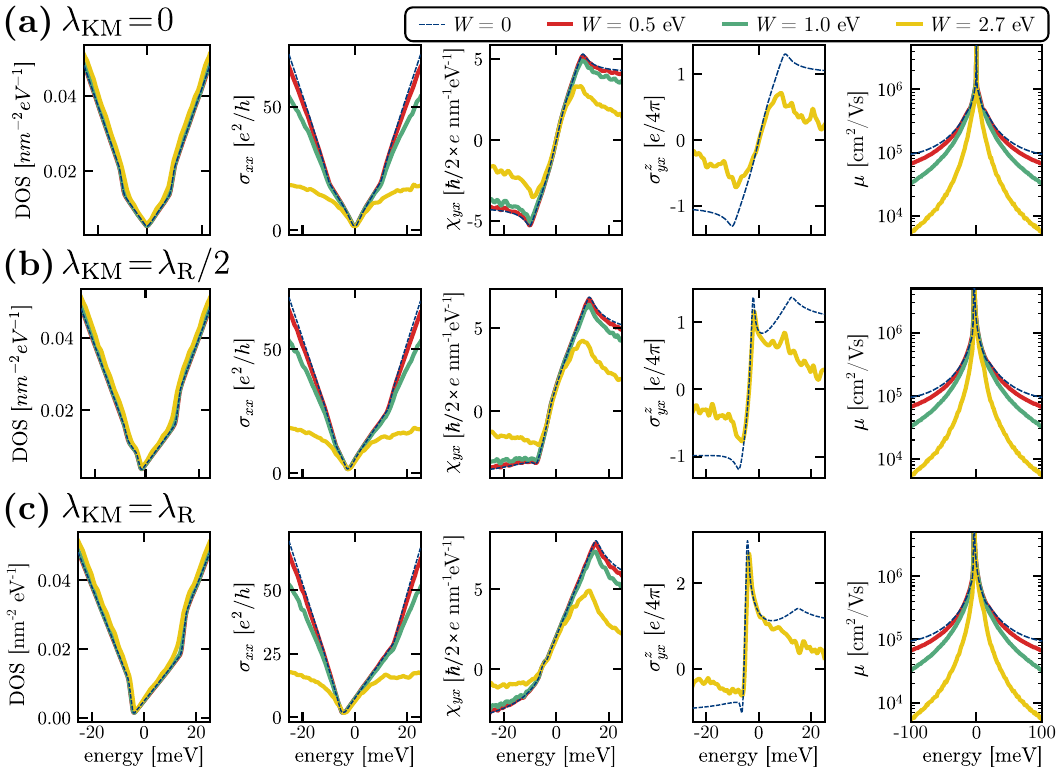}
    \caption{Results of KPM simulations of disordered systems (density of states, longitudinal conductivity, REE spin density, and SHE conductivity) and charge mobility. [$\lambda_\text{R} = 10 \,\text{meV}$]}
    \label{fig:simulations_all}
\end{figure}

Fig. \ref{fig:simulations_all} shows the results for the real-space simulations for $\lambda_\text{KM} / \lambda_\text{R} = 0, \frac{1}{2} \text{ and } 1$ including disorder. These results were obtained in a system of $529$ million unit cells by the KPM method, using a Chebyshev expansion leading to a numerical Jackson broadening of $0.5 \,\text{meV}$ at the band centre. We observe a decrease in the relaxation time $\tau$ due to the effects of disorder scattering, as observed by the decrease of the Fermi surface responses $\sigma_{xx}$ and $\chi_{yx}$; while the REE efficiency remains largely robust. We also include the electron mobility calculations which allow to compare with experimental graphene samples, showing that the disorder values here explored compare reasonably to real graphene samples of medium to good quality \cite{foa-torres_introduction_2020, tan_measurement_2007}. Here, a charge density of $10^{12} \,\text{cm}^{-2}$ is achieved at a Fermi level of approximately $\pm 100 \,\text{meV}$ with respect to charge neutrality.

